\documentclass[reprint,aps,prl,nofootinbib,superscriptaddress]{revtex4-2}

\usepackage{comment}
\usepackage{Journal_names}
\usepackage{booktabs,chemformula}
\usepackage{amssymb}
\usepackage{xspace}
\usepackage{nicefrac}

\newcommand{\multistar}{{\tt multistar}\xspace}
\newcommand{\phaseflow}{{\tt PHASEFLOW}\xspace}
\newcommand{\agama}{{\tt AGAMA}\xspace}
\newcommand{\pc}{\ensuremath{\mathrm{\ pc}}}

\newcommand{\dd}{\mathrm{d}}
\newcommand{\Gyr}{\mathrm{Gyr}}

\newcommand{\mSMBH}{\ensuremath{M_{\rm BH}}}
\newcommand{\RSMBH}{\ensuremath{R_{\rm BH}}}
\newcommand{\mEMRI}{\ensuremath{M_{*}}}
\newcommand{\mDM}{\ensuremath{m_{\rm DM}}}

\newcommand{\rpLISA}{\ensuremath{\mathfrak{r}_{\rm p, LISA}}}
\newcommand{\rpEMRI}{\ensuremath{\mathfrak{r}_{\rm p}}}

\newcommand{\rpEMRIz}{\ensuremath{\mathfrak{r}_{\rm p, 0}}}
\newcommand{\rpDM}{\ensuremath{\mathfrak{R}_{\rm p}}}
\newcommand{\raDM}{\ensuremath{\mathfrak{R}_{\rm a}}}
\newcommand{\eEMRI}{\ensuremath{\mathfrak{e}}}

\newcommand{\eEMRIz}{\ensuremath{\mathfrak{e}_0}}
\newcommand{\eDM}{\ensuremath{\mathcal{E}}}
\newcommand{\raEMRI}{\ensuremath{\mathfrak{r}_{\rm a}}}

\newcommand{\Msolar}{\ensuremath{\mathrm{M}_{\odot}}}

\definecolor{cornellGreen}{HTML}{6EB43F}

\usepackage{tikz}
\usepackage[compat=1.1.0]{tikz-feynman}
\usetikzlibrary{arrows.meta,decorations.pathmorphing}
\usepackage{graphicx}
\usepackage{dcolumn}
\usepackage{bm}

\usepackage{ragged2e}

\usepackage[normalem]{ulem}

\usepackage[pdfnewwindow=true,      
   colorlinks=true,    
  linkcolor=blue,     
 citecolor=blue,     
filecolor=blue,  
urlcolor=blue,      
final=true,
]{hyperref}

\begin{document}
\setlength{\abovedisplayskip}{7pt}
\setlength{\belowdisplayskip}{7pt}


\title{The Irreversible Depletion of Dark Matter Spikes by Extreme-Mass-Ratio Inspirals}

\author{Charlie Sharpe}
\email{charles.sharpe@physics.ox.ac.uk}
\affiliation{Department of Physics, University of Oxford, Parks Road, Oxford OX1 3PU, United Kingdom}

\author{Yonadav Barry Ginat}
\affiliation{Department of Physics, University of Oxford, Parks Road, Oxford OX1 3PU, United Kingdom}

\author{Thomas F.~M.~Spieksma}
\affiliation{Department of Physics, University of Oxford, Parks Road, Oxford OX1 3PU, United Kingdom}

\author{Alexander Heger}
\affiliation{School of Physics and Astronomy, Monash University, Melbourne, Victoria 3800, Australia}

\author{Bence Kocsis}
\affiliation{Department of Physics, University of Oxford, Parks Road, Oxford OX1 3PU, United Kingdom}
\affiliation{St.~Hugh's College, University of Oxford, St.~Margaret's Road, Oxford OX2 6LE, United Kingdom}
\affiliation{Institute of Physics, Eötvös University, Pázmány P.~s.~1/A, Budapest 1117, Hungary}

\date{\today}

\begin{abstract}
Dense dark matter spikes around massive black holes (MBHs) have been widely predicted to produce gravitational-wave dephasing in extreme-mass-ratio inspirals (EMRIs) that are potentially observable by LISA.  We show that EMRIs, the very objects used to probe these spikes, also destroy them. Combining Fokker-Planck models of nuclear star clusters with post-Newtonian three-body simulations of EMRI–dark matter encounters, we demonstrate that repeated gravitational slingshots irreversibly eject dark matter particles from the cluster's loss cone over gigayear time-scales.  For MBHs at $z = 3$ with masses $\lesssim 10^5 \,\Msolar$, even conservative EMRI rates of $\mathcal{O}(1-10) \, \Gyr^{-1}$ suffice to deplete the spike density by orders of magnitude; for more realistic rates of $\mathcal{O}(100-300) \, \Gyr^{-1}$, the mass range extends to $\lesssim 10^6 \,\Msolar$.  As a result, the gravitational-wave dephasing expected from such systems is greatly reduced, narrowing the prospects of detecting DM through this channel.  Only heavy, high-redshift systems with sufficiently low EMRI rates -- for example, a $\sim 10^6 \,\Msolar$ MBH at $z \gtrsim 5$ with $\lesssim 100$ EMRIs per Gyr -- can induce dephasing that exceeds one radian.
\end{abstract}

\maketitle


The origin and nature of dark matter (DM) is a central open problem in physics.  One promising observational window exploits the extreme DM densities that may accumulate around adiabatically grown massive black holes (MBHs)\footnote{By ``massive black hole'' we mean any BH with mass greater than $10^4 \, \Msolar$.} in the early universe, before star formation, where steep density DM ``spikes'' may form \cite{Gondolo_1999} with density profile $\rho_{\rm DM}(r) \propto r^{-\gamma_{\rm sp}}$ and $\gamma_{\rm sp} \in [2.25,2.5]$, implying extreme DM densities in the vicinity of the MBH.  Such spikes have attracted considerable attention because they can imprint measurable phase shifts on gravitational waves (GWs) emitted by intermediate and extreme mass-ratio inspirals (IMRIs and EMRIs) \cite{Eda_2013, Sadeghian_2013,Eda_2015, Yue_2018, Yue_2019a, Yue_2019b, Hannuksela_2020, Kavanagh_2020,Cole_2022, Dai_2022, Becker_2022, Coogan_2022, Speeney_2022, Ghoshal_2023, Dosopoulou_2024, Montalvo_2024, Fischer_2024, Mukherjee_2024, Kavanagh_2024, Alonso_Alvarez_2024, Karydas_2024, Tiruvaskar_2026, Mitra_2025,Vicente_2025, Acevedo_2025, Karydas_2025, Chen_2025, Speri_2026, Chen_2026}. Predicted dephasings of up to $10^2$--$10^6$ radians over multi-year Laser Interferometer Space Antenna (LISA) observations \cite{Colpi_2024} far exceed the $\mathcal{O}(1)$ detection threshold \cite{Hinderer_2008}, motivating extensive study.
The robustness of these predictions, however, is not assured. One must assume that the pre-existing NFW profile is fully formed before the MBH's adiabatic growth, otherwise much shallower $\sim r^{-3/2}$ spike profiles can result \citep{Le_Delliou_2010}. Equal-mass MBH mergers can disrupt or even destroy spikes \cite{Ullio_2001, Milosavljevi_2001, Merritt_2002, Kavanagh_2018}, restricting viable hosts to MBHs with masses $\lesssim 10^6$--$10^7\,\Msolar$, for these are unlikely to have undergone major mergers \cite{Conselice_2006, Fakhouri_2010, Ravi_2015, Rodriguez_Gomez_2016, OLeary_2021}.  Two-body stellar relaxation around a single MBH drives the spike's outer profile towards the much shallower $\rho \propto r^{-\nicefrac32}$ Bahcall-Wolf cusp \cite{Bahcall_1976, Bahcall_1977, Merritt_2004, Gnedin_2004, Merritt_2006, Merritt_2007, Vasiliev_2008, Shapiro_2022,Sharpe_2026}.  Despite these known limitations, many studies treat the spike slope as a free parameter $1 \leq \gamma_{\rm sp} \leq 3$ and fix the density normalisation at the outer boundary of the sphere of influence (SoI) --- the region where the enclosed stellar mass equals the mass of the central MBH --- to $(10^{2}-10^{3})\,\Msolar\pc^{-3}$, resulting in $10$--$15$ orders-of-magnitude variation in DM densities within the innermost few tens of Schwarzschild radii --- precisely the region probed by EMRIs in the sensitivity band of next-generation GW detectors such as LISA \cite{Babak_2017,LISA:2022kgy}, TianQin \citep{TianQin:2015yph}, and DECIGO \citep{Kawamura:2020pcg}.
In this \textit{Letter}, together with a companion paper \cite{Sharpe_2026}, we identify a previously overlooked feedback mechanism: the ejection of DM from the inner regions through slingshot interactions with EMRIs.  Since collisionless DM cannot re-populate the depleted phase space efficiently, due to self-relaxation time-scales being $\sim 10^{70} \, \mathrm{yr}$, this depletion is irreversible. The very objects invoked to detect DM spikes simultaneously destroy them. 

We quantify the spike's depletion by self-consistently combining one-dimensional, orbit-averaged Fokker-Planck simulations of the central few parsecs of nuclear star clusters --- which determine the EMRI flux into the ``loss cone'' (the region where the GW inspiral time-scale is shorter than the two-body angular-momentum diffusion time-scale) --- with post-Newtonian three-body simulations of DM ejection. By studying the extent of DM depletion within the central $\sim 10^3$ Schwarzschild radii of MBHs, we explore how dramatically this narrows the range of cosmological redshifts, MBH masses, and EMRI rates where a DM spike can be detected through GW dephasing. We assume that the initial collisionless spike is smooth, isotropic, and phase mixed, with a thermal eccentricity distribution. Our companion paper \citep{Sharpe_2026} introduces the detailed theoretical framework underlying the results reported here and provides an elaborate discussion of Fokker-Planck models and their validity. Here, we complement that work by directly addressing DM detection prospects over cosmic time through EMRI GW phase shift measurements while accounting for the expected distribution of DM eccentricities.

\paragraph{\textbf{Loss-Cone Boundary Conditions.}}
To determine the distribution of inspiralling stellar-mass BHs (sBHs) in the loss cone, we evolve the stellar distribution outside the loss cone and compute the steady flux of stellar-mass black holes across the loss-cone boundary. The large-scale evolution of the surrounding nuclear star cluster can be modelled with the Fokker-Planck (FP) formalism, which describes the evolution of a system whose dynamics are governed by the cumulative effect of many distant, weak two-body encounters. This treatment is appropriate throughout most of the SoI, but breaks down inside the loss cone, due to the paucity of stars and compact objects there. Instead, the dynamics there are governed by strong encounters with BHs, which FP models completely neglect. Since the detectable GWs from EMRIs probe precisely this inner region, we use the FP evolution only to determine the distribution of EMRI initial conditions at the start of the gravitational-wave inspiral, i.e., at the loss-cone boundary, and treat the subsequent interior dynamics with direct three-body simulations.
Crucially, although strong encounters among stars and compact objects have little effect on the dynamics within the loss cone \cite{Babak_2017, Zhong_2023, Amaro_Seoane_2025}, the same is not true for DM--sBH encounters: DM particles are individually very light and their number density is extremely high, so their evolution is driven by discrete encounters with individual sBHs, which typically transfer energy pushing them to larger radii \cite{Kavanagh_2020, Coogan_2022, Mukherjee_2024, Kavanagh_2024, Alonso_Alvarez_2024, Karydas_2024, Tiruvaskar_2026}.  Moreover, the irreversibility of any depletion by strong EMRI encounters makes the inner spike especially vulnerable to repeated slingshot scatterings over cosmic time.
We restrict our attention to MBHs with masses $10^4$--$10^7\, \Msolar$.  At lower masses, the formation and demographics are uncertain \cite{Portegies_Zwart_2004, Mezcua_2017}, whereas at higher masses, frequent major mergers \cite{Conselice_2006, Fakhouri_2010, Ravi_2015, Rodriguez_Gomez_2016, OLeary_2021} are likely to have disrupted any pre-existing spike \cite{Ullio_2001, Milosavljevi_2001, Merritt_2002, Kavanagh_2018}. 
To determine the rate and orbital distribution with which stellar-mass black holes enter the loss cone and become EMRIs, we solve the orbit-averaged, isotropic, one-dimensional FP equation in energy space, assuming a relaxed thermal eccentricity distribution, i.e., $f_\mathrm{e}(e) = 2e$. We take the nuclear star cluster to contain a central MBH of mass $\mSMBH$, and we evolve the stellar distribution with the publicly available code \phaseflow \citep{Vasiliev_2017}, which is part of the \agama package \citep{Vasiliev_2018}. We adopt a Hernquist profile for the initial stellar density and a Kroupa initial mass function \citep{Kroupa_2001}, discretizing the stellar component into ten mass bins between $0.1\,\Msolar$ and $10\,\Msolar$ to capture the enhanced relaxation associated with mass segregation. We provide a brief description of the FP formulation in the End Matter, but readers are also invited to peruse the companion paper \citep[][\S II]{Sharpe_2026} for further details.
As an aside, in the companion paper \citep[][\S II]{Sharpe_2026} we show that this multi-mass treatment is important due to the way in which FP evolution reshapes steep DM spikes. Two-body relaxation drives the profile towards the familiar $\rho(r) \propto r^{-\nicefrac32}$ Bahcall-Wolf cusp \cite{Bahcall_1976, Bahcall_1977} on a time-scale of order the relaxation time --- $\mathcal{O}(0.1-1) \, \Gyr$ \cite{Merritt_2004, Gnedin_2004, Merritt_2006, Merritt_2007, Vasiliev_2008, Shapiro_2022} --- and, within $2\, \Gyr$, mass segregation shortens this time-scale by one or two orders of magnitude relative to single-mass models \citep[][\S{II.D}]{Sharpe_2026}. Consequently, the DM density within the SoI decreases by several orders of magnitude on gigayear time-scales.
We reiterate that our FP model does not directly evolve EMRIs inside the loss cone, but gives the sBH flux across it. This flux is represented by an energy-dependent sink term and ultimately provides the initial conditions for our three-body calculations.
We assume that the mass of the lighter EMRI component is always $10\, \Msolar$.  This represents a conservative lower bound on the depletion of DM since the true EMRI mass distribution is likely weighted to somewhat larger masses \cite{Barack_2004, Hopman_2009, Aharon_2016, Babak_2017}, leading to a stronger depletion effect. Since theoretical EMRI rates are uncertain, spanning roughly $1$--$1000 \, \Gyr^{-1}$ per MBH \cite{Hils_1995, Sigurdsson_1997, Ivanov_2002, Hopman_2005, Amaro_Seoane_2011, Aharon_2016, Babak_2017,Amaro_Seoane_2018,Panamarev_2019,V_zquez_Aceves_2021,Naoz_2022, Rom_2024}, we keep the rate as a free parameter and use the FP calculation to only determine the normalised distribution of EMRIs entering the loss cone. Most estimates, however, cluster around $100$--$300\, \Gyr^{-1}$.

\paragraph{\textbf{DM Depletion inside the Loss Cone.}}

We now turn to the evolution of DM inside the loss cone, where discrete encounters with inspiralling sBHs dominate the spike's dynamics.  Our aim is to determine the fraction of DM that survives after repeated interactions with the full population of EMRIs experienced by a given MBH. We remark that the total DM and stellar mass enclosed within the loss cone is much smaller than $\mSMBH$, whence apsidal precession due to GR dominates over that due to the Newtonian mass profile \citep[][\S III.B]{Sharpe_2026}, and DM particles evolve independently of one another.  Using a first-principles approach, we employ direct three-body MBH-sBH-DM encounter simulations to determine the probability that a single DM particle is either ejected by a gravitational slingshot or swallowed by one of the BHs during a single EMRI orbit, and then aggregate this probability over a full EMRI inspiral, and over all EMRIs during the MBH's history. This yields the survival probability of a DM particle on a given orbit, and hence the fraction of DM that remains.  Throughout, we work in the hierarchical limit $q = \mEMRI/\mSMBH \ll 1$, where $\mEMRI = 10\,\Msolar$ is the mass of the lighter EMRI component, so that the sBH may be treated as a test particle in the MBH potential, and $\mDM \ll \mEMRI$, where $\mDM$ is the individual DM particle mass, so that the DM particle is a test particle relative to both BHs.
The orbit-averaged probability of a strong encounter between an EMRI and a DM particle scales with its cross section, $\pi b_{\rm strong}^2$, where $b_{\rm strong}$ is the maximum impact parameter for an encounter to be considered strong \cite{Sharpe_2026}. For an encounter in which the change in the DM particle's velocity is of order the velocity itself, the impulse approximation \citep[][Eq.~(3.53)]{Binney_2008} gives $b_{\rm max} \propto q$ with $q = \mEMRI/\mSMBH \ll 1$ the EMRI mass ratio \citep{Sharpe_2026}. Thus the ejection probability may be expressed as
\begin{equation}
    p_{\rm ej} (\rpEMRI, \eEMRI, \rpDM, \eDM, \iota, q) = q^2 A(\rpEMRI, \eEMRI, \rpDM, \eDM, \iota) \,,
    \label{eq:q^2 scaling}
\end{equation}
where $A$ is a dimensionless coefficient that depends on the EMRI periapsis $\rpEMRI$ and eccentricity $\eEMRI$, the DM periapsis $\rpDM$ and eccentricity $\eDM$, and the relative inclination between the DM and EMRI orbits $\iota$. The total number of orbits during an EMRI's inspiral scales as $N_{\rm total} = B(\rpEMRIz, \eEMRIz)/q$ for $q \ll 1$, where $B$ is a proportionality coefficient independent of $q$ \cite[][Appendix F therein]{Sharpe_2026}. Here, $\rpEMRIz$ and $\eEMRIz(\rpEMRIz)$ are the initial EMRI periapsis and eccentricity.  Assuming the ejection probability per EMRI orbit is independent of previous orbits, we multiply these probabilities together to obtain the survival probability, $P_{\rm stay}$, of a DM particle over one complete inspiral
\begin{equation}
    \ln P_{\rm stay}(\rpDM, \eDM) \simeq -q\langle A\rangle_{\rm orbs}B(\rpEMRIz, \eEMRIz)\,,
\end{equation}
where $\langle A\rangle_{\rm orbs}$ denotes the average value of $A$ over all $N_{\rm total}$ orbits.
To obtain the net depletion, we average this ``survival probability'' over the full EMRI population that crosses the loss cone boundary for a given MBH.  After integrating over the unit-normalised periapsis distribution $f_{\rpEMRIz}$ from the FP calculation \cite{Sharpe_2026}, we obtain the fraction of DM remaining on an orbit with periapsis $\rpDM$ and eccentricity $\eDM$ is
\begin{equation}
\begin{aligned}
    \ln f_{\rm rem} (\rpDM, & \,\eDM) = - \langle q\rangle \,N_{\rm EMRI}(z)  \int \dd \rpEMRIz \, f_{\rpEMRIz} \\ 
    & \times \, B(\rpEMRIz, \eEMRIz) \, \langle A \rangle_{{\rm orbs, \, eff}}(\rpEMRIz, \eEMRIz, \rpDM, \eDM)\,, 
\end{aligned}    
\label{eq:ln f_rem}
\end{equation}
where
\begin{align}
    \langle A \rangle_{{\rm orbs, \, eff}} = -\frac{1}{q^2 N_{\rm total}} \ln \left(\int P_{\rm stay} \, f_\iota \, \dd \iota\right) \label{eq:A_orb,eff}
\end{align}
has been averaged over inclination with $f_\iota = \sin \iota/2$, $\eEMRIz = \eEMRIz(\rpEMRIz)$, $\langle q\rangle$ is the average mass ratio, $N_{\rm EMRI}(z) = t(z) \, \Gamma_{\rm EMRI}$ is the total number of EMRIs experienced by redshift $z$ with $t(z)$ the time at $z$, and $\Gamma_{\rm EMRI}$ the EMRI rate.  The derivation of this expression, together with the numerical evaluation of the coefficients entering it, is presented in detail in the companion paper \citep[][\S III.C]{Sharpe_2026}; here we summarise only the ingredients needed to study the detectability prospects.

Crucially, $\ln f_{\rm rem} (\rpDM, \eDM) \propto N_{\rm EMRI} \langle \mEMRI \rangle$ (with $\langle \mEMRI \rangle = \mSMBH \langle q \rangle = 10 \Msolar$ the average sBH mass): the EMRI rate required for a given depletion level scales inversely with the average EMRI mass. For the sake of computational efficiency, we approximate the EMRI distribution at times earlier than the run end time, $t(z)$, to be that at $t(z)$ \citep{Sharpe_2026}.
We determine $\langle A \rangle_{{\rm orbs}, \iota}$ numerically by performing a parameter scan across 8,778 grid points, and interpolating to obtain $A$ across the parameter space. We selected $\eDM = \{$0.05, 0.1, 0.15, $\hdots$, 0.95$\}$, $\rpDM/\RSMBH = \{$5, 6.5, 8.5, 10, 20, 50, 100, 200, 300, 500, $1\mathord,000${}$\}$, $\eEMRI = \{$0, 0.2, 0.4, 0.6, 0.8, 1.0$\}$, and took 7 EMRI periapsis values $\rpEMRI = \rpDM + k(\raDM - \rpDM)$ for $k = \{0, 0.05, 0.25, 0.5, 0.75, 0.95, 1\}$ where $\raDM$ is the DM apoapsis. For each point, we run $10^6$ three-body simulations of a single EMRI with $q = 10^{-2}$ and a single DM particle over one EMRI orbit with the $N$-body code \multistar \citep{multistar_ref}, which includes pairwise 3.5PN corrections.  In each run, the sBH orbit is fixed while the DM orbital angles are isotropically randomised. The DM particle is considered ejected if, at the end of the run, its specific energy in the Schwarzschild spacetime satisfies $E \geq 1$ and its radial velocity is positive. We set $A = 0$ whenever the orbits do not overlap ($\min(\raDM, \raEMRI) < \max(\rpDM, \rpEMRI)$ where $\raEMRI$ is the EMRI apoapsis) or, for computational efficiency, whenever $\rpEMRI < \rpDM < \raEMRI$, where the contributions are small --- including such cases would only strengthen our results.
Finally, we convert the orbit-resolved depletion $f_{\rm rem}$ into the local density reduction $\mathcal{F}_{\rm rem}$. Assuming the remaining DM particles are uniformly distributed in azimuth and mean anomaly, we average over orbital phase yielding the radial probability density $\sigma_{\rm DM}$ of finding a DM particle at a radius $\rpDM \leq r \leq \raDM$ for given $(\rpDM, \eDM)$ \cite{Kocsis_Tremaine2015}. For an initial spike profile of $\rho_{\rm DM}(r) \propto r^{-\gamma}$ and an initial thermal eccentricity distribution $f_{\eDM} = 2\eDM$, the surviving fraction of DM is
\begin{align}
    \mathcal{F}_{\rm rem} = \dfrac{\int \dd \rpDM \dd \eDM \, f_{\rm rem}(\rpDM, \eDM) \, \mathcal{J}(\rpDM, \eDM) \, \sigma_{\rm DM}(\rpDM,\eDM,r)}{\int \dd \rpDM \, \dd \eDM \, \mathcal{J}(\rpDM, \eDM) \, \sigma_{\rm DM}(\rpDM, \eDM, r)} \,, \label{eq:mathcal_F_rem}
\end{align}
with $\mathcal{J}(\rpDM, \eDM) = \eDM \, \rpDM^{2 - \gamma} (1 - \eDM)^{\gamma - 3}$, where we have used that the density of states for a Kepler potential scales as $E^{-5/2}$ (\citep[cf.][Eq.~(3)]{Sharpe_2026}) and $\rho_{\rm DM}(r) \propto r^{-\gamma}$ implies $f(E) \propto E^{\gamma - 3/2}$ \citep{Binney_2008}. This quantity directly measures the suppression of the inner spike relative to its initial profile. 

\paragraph{\textbf{DM Depletion Rates.}}

\begin{figure}
    \centering
    {\includegraphics[width=0.49\textwidth]{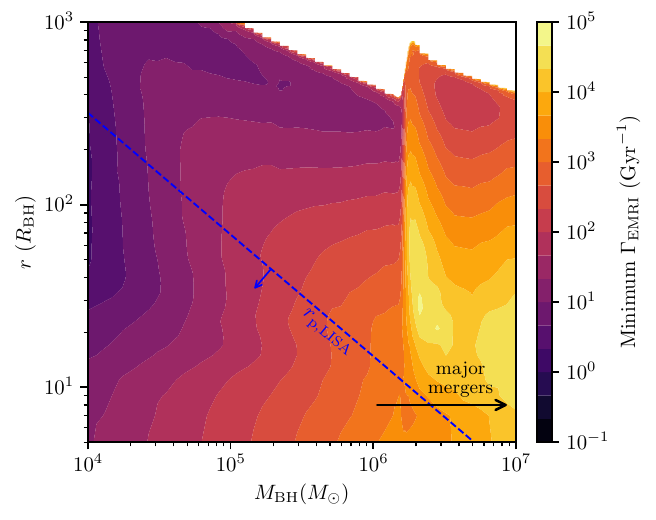}}
     \caption{
     The minimum EMRI rate required for the DM density to be depleted by $\mathcal{F}_{\rm rem} = 10^{-6}$ by $z = 3 \, (t = 2.14 \, \Gyr)$ as a function of MBH mass and distance $r$ from the MBH in units of $\RSMBH$. We assume the lighter component of all EMRIs has a mass of $10 \, \Msolar$.  The blue arrow and dashed line mark $\rpLISA$ (Eq.~\ref{eq:rp_LISA}).
     }
    \label{fig:indiv_min_ER_Rp_MS_z_3}
\end{figure}

Here we combine the orbit-averaged Fokker-Planck calculations with the three-body scattering framework described above to yield the depletion fractions presented below; the numerical implementation and validation are described in detail in Ref.~\citep[][\S III]{Sharpe_2026}. Figure~\ref{fig:indiv_min_ER_Rp_MS_z_3} shows the minimum EMRI rate required to achieve a DM density reduction of $\mathcal{F}_{\rm rem} = 10^{-6}$ (arbitrarily chosen to illustrate the parameter dependence) by $z = 3$, the approximate maximum redshift to which LISA may detect EMRIs \citep{Babak_2017, Chapman_2025, Speri_2026}, as a function of MBH mass and distance $r$ from the MBH, in units of $\RSMBH$, the MBH Schwarzschild radius. A redshift of $z = 3$ corresponds to $t = 2.14 \, \Gyr$ assuming a \emph{Planck} cosmology \citep{Planck:2018vyg}. We have also overlaid $\rpLISA$, the maximum periapsis distance of an EMRI such that its peak GW frequency falls within the LISA band \citep[][Eq.~(36)]{Wen_2003}:
\begin{equation}
    \frac{\rpLISA}{\RSMBH} \approx 37 \left(\frac{1}{1+\eEMRI}\right)^{0.2} \left(\frac{10^{-4} \, \mathrm{Hz}}{f_{\rm min}}\right)^{\frac{2}{3}}\left(\frac{10^6 \, \Msolar}{\mSMBH(1+z)}\right)^{\frac{2}{3}} , \label{eq:rp_LISA}
\end{equation}
where $f_{\rm min} \sim 10^{-4}$ Hz is the lower LISA frequency cut-off \citep{Colpi_2024}. Due to the weak eccentricity dependence, we plot this for $\eEMRI = 0$.
A sharp transition appears at $\mSMBH \approx 1.7 \times 10^6 \Msolar$.  Below this mass, nuclear star clusters have relaxed within $2.14 \, \Gyr$: the stars follow a Bahcall–Wolf cusp and the initial Hernquist stellar profile is ``forgotten''. The EMRI distribution, and hence DM depletion, within such systems is independent of the initial profile. Above this mass, the relaxation time-scale exceeds $2.14 \, \Gyr$ and memory of the initial profile is retained, leading to qualitatively different EMRI distributions and hence different required rates. Since nuclear star cluster initial conditions are poorly constrained, our results above this threshold carry greater uncertainty.
\begin{figure}
    \centering
    {\includegraphics[width=0.49\textwidth]{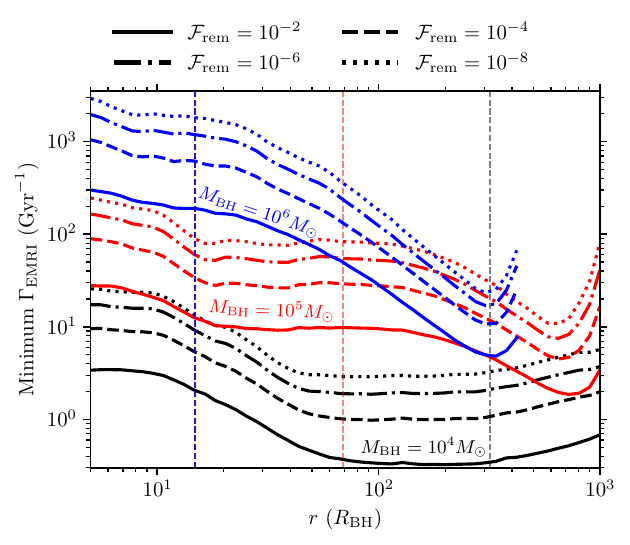}}
     \caption{
     Minimum EMRI rate required to deplete the DM by a factor of $\mathcal{F}_{\rm rem} = \{10^{-8}, 10^{-6}, 10^{-4}, 10^{-2}\}$ for $\mSMBH = \{10^{4}, 10^{5}, 10^{6}\} \, \Msolar$ by $z = 3$. The dashed vertical coloured lines indicate $\rpLISA$ (Eq.~\ref{eq:rp_LISA}).
     }
    \label{fig:min_EMRI_rates_f_rem_z_3}
\end{figure}

\begin{figure}
    \centering
    {\includegraphics[width=0.49\textwidth]{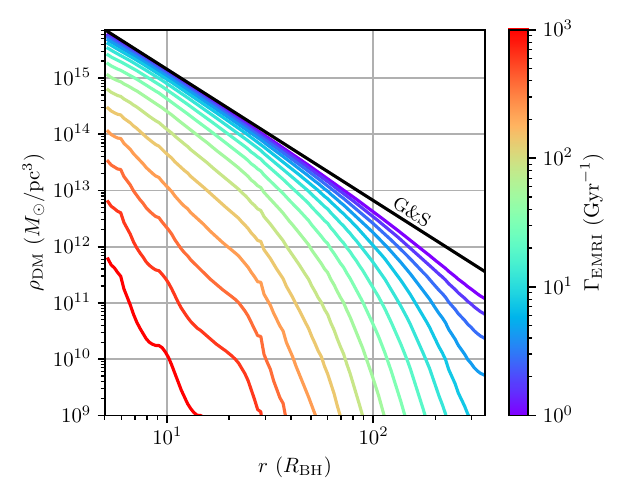}}
     \caption{
     The DM density profile of an EMRI-depleted DM spike by $z = 3$ (2.14 Gyr) around an MBH with mass $\mSMBH = 10^6 \Msolar$ as a function of EMRI rate, $\Gamma_{\rm EMRI}$.  The black line shows the original, undepleted $r^{-7/3}$ Gondolo--Silk profile.
     }
    \label{fig:DM_Density_Profile_z_3}
\end{figure}

We now quantify the EMRI rates required to achieve various DM depletion fractions, $\mathcal{F}_{\rm rem}$.  Figure~\ref{fig:min_EMRI_rates_f_rem_z_3} shows the minimum EMRI rate required for $\mathcal{F}_{\rm rem} = \{10^{-8}, 10^{-6}, 10^{-4}, 10^{-2}\}$, and $\mSMBH = \{10^4, 10^5, 10^6\} \, \Msolar$ by $z = 3$. Depletion is efficient across a wide range of masses and radii, and is greatest for small MBH masses. We illustrate the impact of EMRI-induced depletion on the DM density profile in Figure~\ref{fig:DM_Density_Profile_z_3}. It shows the case $\mSMBH = 10^6 \Msolar$ for different values of $\Gamma_{\rm EMRI}$. For a canonical Gondolo--Silk $r^{-7/3}$ spike formed from an initial NFW profile, substantial evaporation occurs. Lighter MBHs would exhibit even more significant depletion for a given $\Gamma_{\rm EMRI}$. At radii larger than that shown here --- outside of the loss cone --- the density is set by our Fokker-Planck model.

The bottleneck of EMRI-induced DM depletion lies in the difficulty of removing low-eccentricity DM particles \citep{Sharpe_2026}. These orbits span a narrower radial range, reducing the overlap with EMRIs, thereby leading to fewer DM ejection opportunities and hence less depletion. Thus, we expect spikes that have been depleted due to EMRIs to almost exclusively consist of low-eccentricity DM particles. Conversely, DM spikes dominated by low (high) eccentricity DM particles will undergo weaker (stronger) subsequent depletion.
Figures~\ref{fig:indiv_min_ER_Rp_MS_z_3}--\ref{fig:DM_Density_Profile_z_3} show that EMRI-driven depletion is efficient across much of the relevant mass range for $z = 3$. Optimistic EMRI rates of $\sim 10^3 \, \Gyr^{-1}$ \citep{Sigurdsson_1997, Amaro_Seoane_2018, V_zquez_Aceves_2021} deplete the inner spike by many orders of magnitude for MBHs of mass $\mSMBH \lesssim 10^6 \Msolar$. Even conservative rates of $\sim (1-10) \, \Gyr^{-1}$ achieve this for $\mSMBH \lesssim 10^5 \Msolar$.  At larger masses, such as $\mSMBH = 10^7\Msolar$, the depletion is less efficient --- yet those BHs are expected to have undergone major mergers at some point in their lifetime \cite{Conselice_2006, Fakhouri_2010, Ravi_2015, Rodriguez_Gomez_2016, OLeary_2021}, which directly disrupt the DM spike anyway \cite{Ullio_2001, Milosavljevi_2001, Merritt_2002, Kavanagh_2018}. Thus, over a broad and astrophysically relevant parameter range, EMRIs alone strongly suppress the DM densities within the loss cone.
%

\paragraph{\textbf{GW Dephasing in Depleted Spikes.}} 
One of the main detection channels of collisionless DM spikes is the GW dephasing they induce in EMRIs within the LISA band, $10^{-4}$--$10^{-1}$ Hz \cite{Amaro_Seoane_2017}. We now estimate the extent to which this signal is reduced once EMRI-driven depletion is taken into account.
Assuming quasi-circular evolution, the total GW phase accumulated as the EMRI's orbital frequency evolves from $f_{\rm start}$ to $f_{\rm end}$ is \cite{Mitra_2025}
\begin{align}
    \Phi_{\rm total} = 4 \pi \int_{f_{\rm start}}^{f_{\rm end}} \frac{f'}{\dot{E}_{\rm DF} + \dot{E}_{\rm GW}} \frac{\dd E_{\rm orbit}\left(f'\right)}{\dd f'} \dd f',
\end{align}
where we only account for energy losses due to GW emission and dynamical friction, neglecting, for example, accretion \citep{Karydas_2024} and relativistic corrections \citep{Karydas_2025}. Here, $E_{\rm orbit}$ is the orbital energy, $\dot{E}_{\rm GW} = -[32{\rm G}^{7/3}/(5{\rm c}^5)] (2 \pi f M_{\rm T})^{10/3} \eta^2$ is the leading-order GW energy dissipation rate \cite{Peters_1963}, with $\eta$ being the symmetric mass ratio, $M_{\rm T} = \mSMBH + \mEMRI$ the total mass, and $\dot{E}_{\rm DF} = -4\pi {\rm G}^2 M_{\rm T}^2 \rho(r) \eta^2 \ln \Lambda (1+v^2/{\rm c}^2)^2/(v(1-v^2/{\rm c}^2))$ the relativistically corrected energy dissipation due to dynamical friction \cite{Chandrasekhar_1943,Kavanagh_2024, Mitra_2025} with $v = \sqrt{GM_{\rm T}/r}$ the EMRI's orbital velocity. The DM-induced dephasing, $\Delta \Phi$, is obtained by subtracting the phase accumulated in the presence of a DM spike from that in vacuum, where $\dot{E}_{\rm DF} = 0$. LISA is expected to be sensitive to a minimum dephasing of $\mathcal{O}(1)$ radians or more over a 5-year observation run \cite{Hinderer_2008}.
\begin{figure}
    \centering
    {\includegraphics[width=0.49\textwidth]{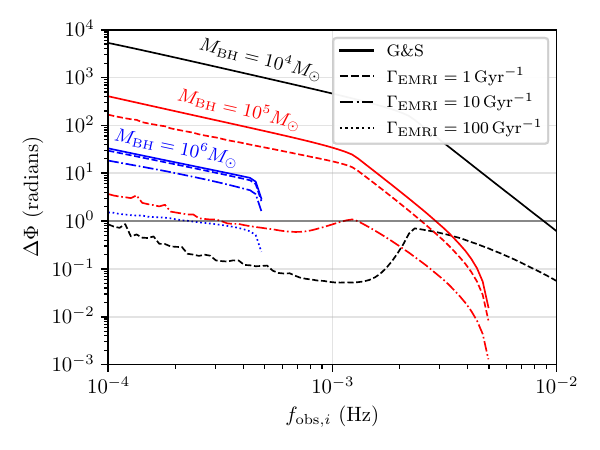}}
     \caption{
     DM-induced GW dephasing over a $5\,$yr observation with initial observed GW frequency $f_{{\rm obs},i}$ for the undepleted $r^{-7/3}$ Gondolo--Silk spike compared to an EMRI-depleted spike at $z = 3$ for EMRI rates of $\Gamma_{\rm EMRI} \in \{1, 10, 100\} \, \Gyr^{-1}$ and MBH masses $\mSMBH \in \{10^4, 10^5, 10^6\}\, \Msolar$ in black, red, and blue respectively.  The darker line at $\Delta \Phi = 1$ corresponds to the approximate LISA detection threshold.
     }
    \label{fig:dephasing_plot_5_z_3}
\end{figure}
Figure~\ref{fig:dephasing_plot_5_z_3} shows the dephasing in the detector frame for a circular EMRI inside an EMRI-depleted spike, as well as an undepleted Gondolo--Silk \cite{Gondolo_1999} spike, whose initial observed GW frequency is $f_{{\rm obs}, i}$ (frequency of $(1+z) f_{{\rm obs}, i}$ in the EMRI frame) for $z = 3$. We integrate until either $5\,$yr have passed in the detector frame (corresponding to $5/(1+z)$ years in the EMRI frame), or until the EMRI's semi-major axis falls below $5\,\RSMBH$ as we have not modelled radii smaller than this. For the lightest systems we consider, $\mSMBH = 10^4\, \Msolar$, an EMRI rate of $\mathcal{O}(1) \, \Gyr^{-1}$ is sufficient to suppress $\Delta \Phi$ to below unity. For $\mSMBH = 10^6\,\Msolar$, the corresponding threshold rises to $\mathcal{O}(100) \, \Gyr^{-1}$. We emphasise that most EMRI rate predictions lie around $(100 - 300) \, \Gyr^{-1}$ \citep{Amaro_Seoane_2018, Amaro_Seoane_2011, Aharon_2016, Panamarev_2019, V_zquez_Aceves_2021, Naoz_2022, Rom_2024}.

The precise detectability threshold depends on factors such as the signal-to-noise ratio, waveform systematics, observation time, and the detector itself. Overall, however, our results suggest that detecting a DM spike with LISA at $z = 3$ through EMRI dephasing will likely require a heavy host MBH with low EMRI rates: for example, $\mSMBH = 10^5\, \Msolar$ or $10^6 \,\Msolar$ requires $\Gamma_{\rm EMRI} \lesssim 10\, \mathrm{ Gyr}^{-1}$ or $\Gamma_{\rm EMRI} \lesssim 100\,\mathrm{Gyr}^{-1}$, respectively.

\begin{figure}
    \centering
    {\includegraphics[width=0.49\textwidth]{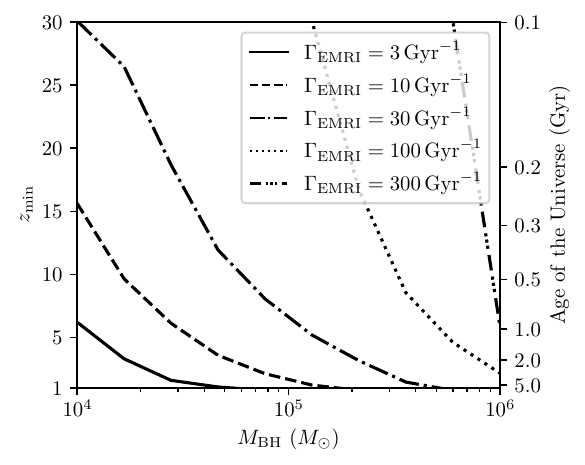}}
     \caption{
     The minimum redshift, $z_{\rm min}$, such that there exists at least one frequency in the LISA band where an EMRI will undergo dephasing greater than 1 radian --- the approximate detection threshold --- over a $5\,$yr observation as a function of $\mSMBH$ and $\Gamma_{\rm EMRI}$.  We assume the lighter EMRI component has mass $10 \, \Msolar$.
     }
    \label{fig:1_rad_plot}
\end{figure}

Notably, higher redshift (younger) systems may be less depleted as they have (1) experienced fewer EMRIs in their lifetime (for a given EMRI rate) and (2) had less time to relax and undergo mass segregation, reducing the accumulation of sBHs with small semi-major axes (\citep[][\S III.B]{Sharpe_2026}).  Figure \ref{fig:1_rad_plot} shows the minimum redshift, $z_{\rm min}$, for which $\Delta \Phi \geq 1$ radian over a $5\,$yr observation within the LISA band. A spike at $z = 30$ can withstand EMRI rates a factor of $\mathcal{O}(10)$ greater than a spike at $z = 3$ before becoming undetectable.  Detecting DM spikes with EMRIs therefore favours higher-redshift sources, motivating instruments with reach beyond that of LISA. We note that our dephasing predictions for higher redshift systems, which have not relaxed, are more uncertain due to stronger dependence on the initial conditions of the nuclear star cluster, which are poorly constrained.

%
 
%
\paragraph{\textbf{Conclusions.}}
%
In this \emph{Letter}, we address one of the outstanding problems in the study of DM spikes: their survival in realistic astrophysical environments. By treating collisionless spikes around MBHs as dynamical components of the surrounding galactic nuclei, we have shown that gravitational slingshot ejections by EMRIs significantly and irreversibly deplete DM spikes. As a result, the canonical $r^{-7/3}$ Gondolo-Silk spike does not represent a long-lived equilibrium solution around MBHs with mass $\mSMBH \lesssim 10^6\,\Msolar$ at $z = 3$. 
This effect limits the possibility of DM detection using EMRI GW dephasing measurements to much higher redshifts (Figure~\ref{fig:1_rad_plot}), requiring future detectors beyond LISA.
%

%
Several caveats apply. We assume that all MBHs are embedded in nuclear star clusters, which is not guaranteed, particularly at higher masses \citep{Neumayer_2020, Hoyer_2021}. We approximated the final EMRI distribution from our FP models to represent the distribution at all earlier times; we assume that the DM ejection probability per EMRI orbit is independent of previous orbits%
\footnote{
\citet{Karydas_2026} posted to arXiv while this manuscript was under review. They study the same physical system but instead use a Fokker-Planck-type approach aimed at modelling the effect of both weak \textit{and} strong interactions. They argue that their results differ from ours and that our assumption of independent ejection probabilities per orbit causes us to significantly overpredict the extent of depletion. We validate this assumption, however, and show that our results are consistent with complete inspiral $N$-body simulations in the companion paper \citep[][Appendix G]{Sharpe_2026}.
}%
; and we note that results could differ significantly for self-interacting DM models due to efficient refilling of the depleted phase space.
While we focus on the impact of EMRI-induced DM depletion on GW dephasing, the implications extend more broadly. Spikes have been invoked to constrain DM properties through their influence on S-star orbits around Sgr~A$^*$ \cite{Nampalliwar_2021, Shen_2023, Gustafson_2025}, through $\gamma$-ray and other electromagnetic signatures of DM annihilation or decay \citep{Bergstrom_2012, Bertone_2006, Aharonian_2008, Bertone_2009, Wanders_2015, Alvarez_2021, Freese_2022, Balaji_2023, Bertone_2024}, and through non-gravitational DM--Standard-Model interactions \cite{Shapiro_2016, Fujiwara_2024, Acevedo_2025, Meighen_2025}. DM spikes also affect the post-merger BH ringdown signal \cite{Dong_2025}. On larger scales, it was proposed that spikes might alter the evolution of super-massive BH binaries, imprinting signatures on the stochastic nHz GW background observable by Pulsar Timing Arrays \cite{Ghoshal_2023, Hu_2025, Chen_2025, Shen_2025b}, and have been proposed as a resolution to the ``final parsec'' problem \cite{Alonso_Alvarez_2024, Tiruvaskar_2026}. While some of these channels involve collisional or self-interacting DM models that are not directly studied here, our results suggest that any signature relying on large DM densities in the inner spike may be weakened for MBHs in the mass range $10^4$--$10^6\,\Msolar$, where DM densities are suppressed by factors $F_{\rm rem} \sim 10^{-2}$--$10^{-8}$ at $z = 3$, depending on the EMRI rate and distance from the MBH (Figures~\ref{fig:indiv_min_ER_Rp_MS_z_3}--\ref{fig:DM_Density_Profile_z_3}). 
%


\paragraph{\textbf{Acknowledgements.}}

We thank John Magorrian for insightful conversations about strong encounters and direct plunge orbits, Fabio Antonini, Fani Dosopoulou and Johan Samsing for helpful discussions on DM spikes and EMRIs. The \multistar documentation is available online.\footnote{ \url{https://2sn.erc.monash.edu/multistar/doc/index.html}.} This work was supported by the STFC (grant No.~ST/W000903/1), and by a Leverhulme Trust International Professorship Grant (No.~LIP-2020-014). C.S.~acknowledges funding from an STFC studentship. Y.B.G.'s work was partly supported by the Simons Foundation via a Simons Investigator Award to A.A.~Schekochihin. T.F.M.S.~acknowledges support from a Royal Society University Research Fellowship (URF-R1-231065). A.H.\ acknowledges support from the Australian Research Council through grants DP240101786 and DP240103174.


\bibliographystyle{apsrev4-2}
\bibliography{apssamp}

\section*{End Matter}
\appendix*
\label{Appendix}

\paragraph{\textbf{The Fokker-Planck Model.}}

Here, we briefly detail the Fokker-Planck formulation --- further details can be found in the companion paper \citep[][\S II.A]{Sharpe_2026}. The 1D orbit-averaged, isotropic FP equation in energy space reads
\begin{align}
    -\frac{\dd q}{\dd E} \frac{\partial f_s}{\partial t}=\frac{\partial}{\partial E}\left[D_{E, s}(E) f_s(E)+D_{EE, s}(E) \frac{\partial f_s}{\partial E}\right]\,, \label{eq:FP}
\end{align}
where $f_s(E,t)$ is the phase-space distribution function of species $s$, $q(E)$ is phase-space volume enclosed by the orbital energy surface, and $D_{EE,s}(E)$ and $D_{E,s}(E)$ are the energy diffusion and dynamical friction coefficients, respectively. These are given by 
\begin{align}
\begin{aligned}
    q(E) &= \frac{2^{3/2}\pi^3}{3} \mathrm{G}^3 \mSMBH^3 E^{-3/2}\,,\\
    D_{E E, s}(E) &=A_0 \sum_{k} m_k^2\left(\int_E^{\infty} f_k\left(E_k\right) q\left(E_k\right) \mathrm{d} E_k + \right. \\
    & \left.q(E) \int_{-\infty}^E f_k\left(E_k\right) \mathrm{d} E_k\right)\,, \\
    D_{E, s}(E) &=A_0 m_s \sum_{k}  m_k\left(\int_E^{\infty} f_k\left(E_k\right) \frac{\partial q}{\partial E_k} \mathrm{d} E_k\right)\,,\\
    A_0 &= 64 \pi^4 \mathrm{G}^2 \ln \Lambda\,.
\end{aligned} \label{eq:q,D_EE,D_E}
\end{align}
Each sum runs over all species $k$ that interact gravitationally with species $s$, and $\ln \Lambda = 15$ is the Coulomb logarithm \citep{Vasiliev_2017}. This equation describes how the orbital energies $E$ of individual particles (e.g. stars or DM particles) diffuse in time due to distant weak interactions with objects in an isotropic cluster having a relaxed thermal eccentricity distribution, $f_e = 2e$. The flux across the loss-cone boundary is then calculated using an energy-dependent sink term \citep{Vasiliev_2017}
\begin{align}
    \nu(E) = \frac{T_J^{-1}}{\left(u^2 + u^4\right)^{1/4} + \ln(1/\mathcal{R}_{\rm LC})}\,, \label{eq:nu(E)}
\end{align}
where $u = (1/\mathcal{R}_{\rm LC})(P(E)/T_J)$ and $\mathcal{R}_{\rm LC} = L_{\rm LC}^2/L_{\rm c}^2$ with $P(E)$ the period for an orbit with energy $E$, $T_J(E)$ the angular momentum relaxation time-scale, $L_{\rm LC}(E)$ the angular momentum required to enter the loss cone, and $L_{\rm c}(E)$ the angular momentum of a circular orbit with energy $E$.

\end{document}